\documentclass[aps,prl,amsmath,floatfix,amssymb,twocolumn,showpacs,superscriptaddress]{revtex4-1}

\usepackage[dvips]{graphicx}
\usepackage{epsfig}                
\usepackage{colordvi}              
\usepackage{xcolor}

\begin{document}

\preprint{}

\title{Packing Confined Hard Spheres Denser with Adaptive Prism Phases}



\author{Erdal C. O\u{g}uz}
\email[]{ecoguz@thphy.uni-duesseldorf.de}
\homepage[]{http://www2.thphy.uni-duesseldorf.de/~ecoguz/}
\affiliation{Institut f\"ur Theoretische Physik II: Weiche Materie, Heinrich-Heine-Universit\"at D\"usseldorf,
  Universit\"atsstra{\ss}e 1, 40225 D\"usseldorf, Germany}

\author{Matthieu Marechal}
\affiliation{Institut f\"ur Theoretische Physik II: Weiche Materie, Heinrich-Heine-Universit\"at D\"usseldorf,
  Universit\"atsstra{\ss}e 1, 40225 D\"usseldorf, Germany}

\author{Fernando Ramiro-Manzano}
\affiliation{Nanoscience Laboratory, Dept. Physics, University of Trento, Via Sommarive 14, I-38050 Trento, Italy}

\author{Isabelle Rodriguez}
\affiliation{Centro de Tecnologias Fisicas, Unidad Asociada ICMM/CSIC-UPV, Universidad Polit\'ecnica de Valencia,
Av. Los Naranjos s/n, 46022 Valencia, Spain}
\affiliation{Instituto de Ciencia de Materiales de Madrid CSIC, 28049 Madrid, Spain}

\author{Ren\'e Messina}
\affiliation{Institut f\"ur Theoretische Physik II: Weiche Materie, Heinrich-Heine-Universit\"at D\"usseldorf,
  Universit\"atsstra{\ss}e 1, 40225 D\"usseldorf, Germany}
\affiliation{Institut de Chimie, Physique et Mat{\'e}riaux (ICPM), Universit{\'e} de Lorraine,
  1 Bld Arago, 57078 Metz - Cedex 3, France}

\author{Francisco J. Meseguer}
\affiliation{Centro de Tecnologias Fisicas, Unidad Asociada ICMM/CSIC-UPV, Universidad Polit\'ecnica de Valencia,
Av. Los Naranjos s/n, 46022 Valencia, Spain}
\affiliation{Instituto de Ciencia de Materiales de Madrid CSIC, 28049 Madrid, Spain}

\author{Hartmut L\"owen}
\affiliation{Institut f\"ur Theoretische Physik II: Weiche Materie, Heinrich-Heine-Universit\"at D\"usseldorf,
  Universit\"atsstra{\ss}e 1, 40225 D\"usseldorf, Germany}



\date{\today}

\begin{abstract}
We show that hard spheres confined between two parallel hard plates
pack denser with periodic adaptive prismatic structures which are composed of alternating
prisms of spheres. The internal structure of the prisms adapts to the slit height
which results in close packings for a range of plate separations, just above 
the distance where three intersecting square layers fit exactly between the plates.
The adaptive prism phases are also observed in real-space experiments on confined sterically stabilized 
colloids and in Monte Carlo simulations at finite pressure.
\end{abstract}



\maketitle


How to pack the largest number of hard objects in a given volume is a classic optimization 
problem in pure geometry \cite{Hales2006}. The close-packed structures obtained from such optimizations
are also pivotal in understanding the basic physical mechanisms behind freezing \cite{Bernal1960,Stillinger1964}
and glass formation \cite{Zallen1983}. Moreover, close-packed structures are highly relevant to 
numerous applications ranging from packaging macroscopic bodies and granulates \cite{Edwards1994} 
to the self-assembly of colloidal \cite{Russel1989} and biological \cite{Liang2001,Purohit2003} 
soft matter. For the case of hard spheres, Kepler conjectured that the highest-packing density should be 
that of a periodic face-centered-cubic (fcc) lattice composed of stacked hexagonal layers; it took until 2005
for a strict mathematical proof \cite{Hales2005}.
More recent studies on close packing concern either non-spherical hard objects \cite{Torquato_review}
such as ellipsoids \cite{Donev2004,Philipse2007}, convex polyhedra \cite{Damasceno2012,Torquato2009}
(in particular tetrahedra \cite{Haji2009}), and irregular non-convex bodies \cite{Graaf2011}  
or hard spheres confined in hard containers \cite{Pickett2000,Mughal2011,Kegel2005} or 
other complex environments.

\begin{figure}
  \includegraphics[width=6.5cm]{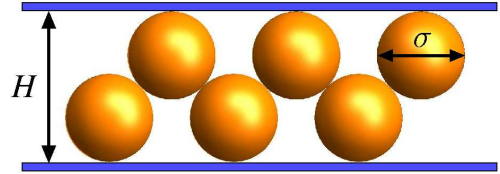} 
  \caption{\label{fig:fig1} Schematic illustration of hard spheres of diameter $\sigma$
    confined between two parallel hard plates of separation $H$.}
\end{figure}

If hard spheres of diameter $\sigma$ are confined between two hard parallel plates of distance 
$H$, as schematically illustrated in Fig.\ \ref{fig:fig1}, the close-packed volume fraction $\phi$ 
and its associated structure depend on the ratio $H/\sigma$. 
Typically, the complexity of the observed phases increases tremendously on confining the system. 
Parallel slices from the fcc bulk crystal are only close-packed for certain 
values of $H/\sigma$: A stack of $n$ hexagonal (square) layers aligned with the walls, 
denoted by $n\triangle$ ($n\square$), is  best-packed at the plate separation  $H_{n\triangle}$ 
($H_{n\square}$) where the layers exactly fit between the walls. 
Clearly, for the minimal plate distance $H \equiv H_{1\triangle} =\sigma$, packing by a hexagonal 
monolayer is optimal. Increasing  $H/\sigma$ up to $H_{2\triangle}$, a buckled monolayer \cite{Pieranski_1983} 
and then a rhombic bilayer \cite{Schmidt1,*Schmidt2} become close-packed. 
However, for $H_{2\triangle}<H<H_{4\triangle}$, the close-packed structures are much more complex and 
still debated. Both, prism phases with alternating parallel prism-like arrays composed of hexagonal 
and square base \cite{Neser1997,Fortini} and morphologies derived from the hexagonal-close-packed (hcp) 
structure \cite{Manzano2007,Fontecha2007} were proposed as possible candidates. 

For confined hard spheres, the knowledge and control  over the close-packed configuration is
of central relevance for at least two reasons: 
First, the hard sphere system away from close-packing is of fundamental interest as a 
quasi-two-dimensional statistical mechanics model. At low densities, a hard sphere gas is stable,
which will crystallize as the density is increased beyond some threshold value. As such, 
the model represents a classical route to understand freezing between two and three spatial 
dimensions \cite{Alsayed2011}. 
The associated fluid--solid transition will be strongly affected by the close-packed structure. 
Second, the confined hard sphere model is almost perfectly realized in nature by mesoscopic
sterically-stabilized colloidal suspensions \cite{Neser1997,Manzano2009} which can be confined between 
glass plates providing a slit-like confinement. At high imposed pressures, colloids will self-assemble 
into the close-packed structures. It has been shown that this is the key for 
the controlled fabrication of nano-sieves and of membranes with desired morphology \cite{Goedel}.

In this Letter, we explore the close-packed structures of 
confined hard spheres by combining numerical optimization, experiments and computer simulation.
Using a systematic penalty optimization method, we find the whole cascade of close-packed structures
in the range of plate distances $H_{1\triangle}<H<H_{4\triangle}$. As an important building block
for close-packing, an {\it adaptive\/} prism is identified which adjusts its internal 
structure flexibly to the slit height $H/\sigma$. This prism has a base with a rhombic symmetry
and neighboring prismatic arrays are shifted relative to each other.
The resulting adaptive structure maximizes the packing fraction in the regime beyond $H_{3\square}$. 
We also propose a further close-packing prism phase of square symmetry that 
packs densest in the regime just beyond $H_{4\square}$ and shows a two-dimensional relative lateral shift
between the prisms. 
We confirm the stability of the new adaptive prismatic structures both in real-space experiments on 
confined sterically stabilized colloids and in Monte Carlo simulations at finite pressure. 
In the following, we first describe the results from the penalty method, then discuss 
real-space data for confined colloidal samples and subsequently turn to Monte Carlo simulation results. 
Details of the numerical methods, experiments and simulations are listed in the 
Supplemental Material~\cite{suppinfo}. 

In our numerical calculations, we consider periodic structures with up to 12 particles per unit 
cell thereby covering all hitherto proposed structures~\cite{suppinfo}. To maximize the
packing fraction $\phi$, we optimized the cell shape and the particle coordinates of these 
structures. However, investigating the dense-packing of hard spheres accommodates a constrained 
optimization: the free volume must be minimized under the constraint of non-overlapping spheres. 
To circumvent the discontinuous, constrained optimization, we employed the \textit{penalty method} 
\cite{Fiacco1968} in our numerical calculations. 
By adding a penalty term in case the spheres intersect which depends continuously on the overlap volume, 
we obtained a continuous and unconstrained penalty function which can be minimized in the classic 
way to predict the optimal particle coordinates.
The penalty method offers the flexibility to use a relatively broad range of candidate crystalline 
lattices and has recently been shown to allow a very efficient handling of packing 
problems \cite{Assoud_penalty}.

\begin{figure}[!ht]
  \includegraphics[width=7.7cm]{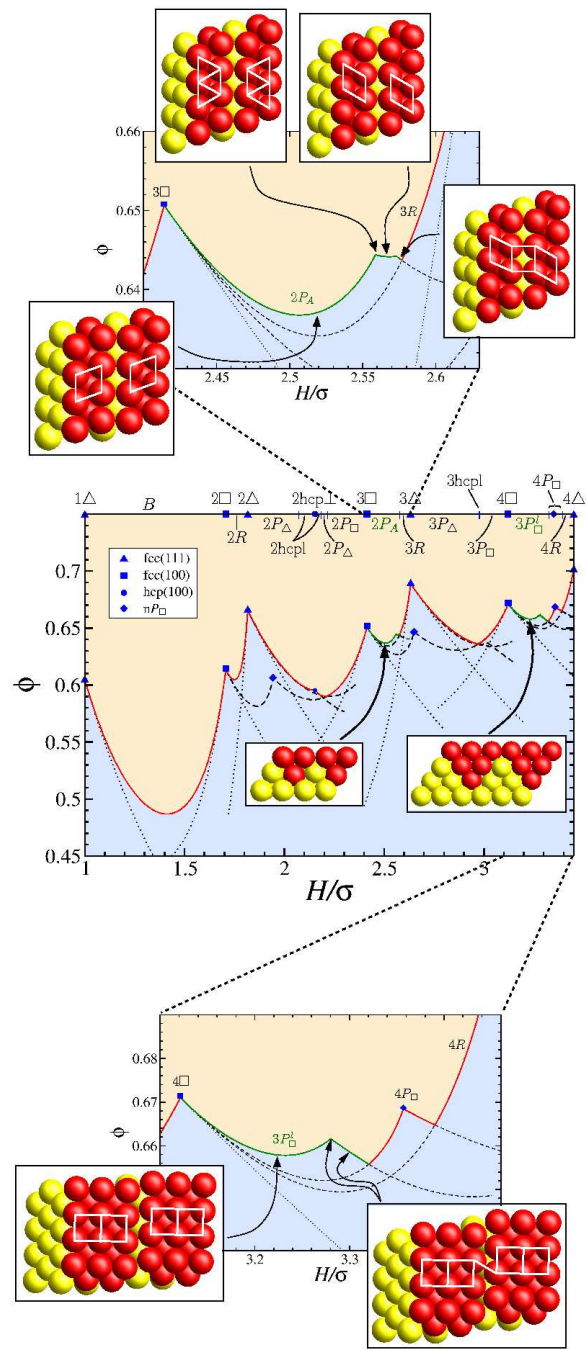} 
  \caption{\label{fig:fig2} Packing fraction $\phi$ versus dimensionless height $H/\sigma$. 
    The best-packing phases are indicated by symbols on the top axis of the middle panel and their 
    packing fractions are shown as the full lines. For clarity, the regions below and above the best 
    packing are colored differently. The new prism phases  are denoted by green lines. 
    Dashed and dotted lines denote the non-close packed $n\square$ [fcc(100)], $n\triangle$ [fcc(111)], 
    $n$hcpl, $n P_\triangle$ and $n P_\square$. The top and bottom panel show enlargements of the regions 
    where the new prism phases were found. Side views (middle panel) and top views (top and bottom panels) 
    show the structure of these phases, where white lines denote bonds between touching particles.
}
\end{figure}

The resulting volume fractions of the densest packed phases are shown in Fig.\ \ref{fig:fig2} as a 
function of $H/\sigma$ in the regime between the hexagonal monolayer $1\triangle$ ($H/\sigma=1$) and 
the triangular tetralayer $4\triangle$ ($H/\sigma=\sqrt{6}+1$). For $H_{1\triangle}<H<H_{2\triangle}$ 
the classic sequence \cite{Pieranski_1983,Schmidt1,*Schmidt2,Grandner_JCP_2008}
 $1\triangle\to B\to 2\square \to 2R \to 2\triangle$ is confirmed. 
Here, $B$ is a buckled hexagonal layer with rectangular symmetry and 
the $2R$ crystal consist of two staggered rhombic layers.

%
%
For $H_{2\triangle}<H<H_{4\triangle}$, there is a much more complex 
cascade of close-packed structures. In the transition regime $n\triangle \to (n+1)\square$ for $n=2,3$, 
on the one hand, we recover all of the phases found previously. 
Here, we obtain the sequence $2\triangle \to 2 P_\triangle \to 2\mathrm{hcpl} 
\to 2\mathrm{hcp}\bot \to 2\mathrm{hcpl} \to 2 P_\square \to 3\square$ and $3\triangle \to 3 P_\triangle 
\to 3\mathrm{hcpl} \to 2 P_\square \to 3\square$, where the following phases are encountered: 
The $2n$-layered phases $nP_{\triangle}$ and $nP_{\square}$ consist of alternating prism-like 
dense-packed $n$-layered arrays of spheres with triangular ($\triangle$) and square ($\square$) 
basis shapes (\cite{Fortini,Neser1997,Manzano2009}). 
Moreover, the $2n$-layered phases $n\mathrm{hcp}\bot$~\footnote{This phase structure corresponds 
to the (100) plane of the hexagonal close-packed solid and, therefore, it is referred to as 
$n\mathrm{hcp}(100)$ in Ref.~\cite{Manzano2007}} and $n$hcp-like with rectangular 
symmetry (see \cite{Fontecha2007,Manzano2007}) are found. For $n=3$, however, 
the $n$hcp-like phase is only close-packed in a tiny regime, whereas $n$hcp$\bot$ is not found at all 
(see Fig.\ \ref{fig:fig2}). 

In the range $3\square \to 3\triangle$,   
the new adaptive prism phase $2P_A$ is predicted to be close-packed. Representative intra-layer 
touching bonds are indicated by white lines in Fig.\ \ref{fig:fig2} (upper and lower panel) 
to underline  the symmetry of the corresponding prismatic structure. 
The $2P_A$ phase adapts its internal structure flexibly to an increase of the slit width $H$. 
In fact, the symmetry of its prism blocks is rhombic which spans the whole range between
the square symmetry of the underlying phase $3\square$  and the triangular base 
(see white lines in Fig.\ \ref{fig:fig2}, upper panel). 
Likewise, we noticed the stability of the $3P^l_{\square}$ 
prism phase with square basis shape (cf.\ Fig.\ \ref{fig:fig2}, 
lower panel) in the transition regime $4\square \to 4\triangle$ whose prisms exhibit 
a longitudinal shift (ie.\ in the lenghtwise direction of the prisms) in addition to the 
usual shift perpendicular to the lengthwise direction of the prisms. 
Finally, the other densest packed phases are multilayered rhombic phases
$3R$ and $4R$ as well as a square prism phase $4P_\square$ see Fig.\ \ref{fig:fig2}.

To verify our theoretical results, we performed real-space experiments with nanometer-sized
colloids. We employed Polystyrene particles with diameters $\sigma$ in the range from 245 nm to 800 nm 
(Ikerlat Polymers) to study a certain of $H/\sigma$ values. 
We created a confining wedge cell with a very small opening angle ($10^{-4}$ rad) and 
slit height $H<6\mu$m (see \cite{suppinfo}). The varying slight height inherent to the wedge geometry 
allows many transitions between different crystals in the same cell. 
Finally, after the sample was dried, we detached the Polystyrene covering plate. Some particles stuck to the 
covering plate during its removal resulting in holes in the top layer of particles, which allowed us 
to study the structure in the layers below. We recorded Scanning Electron Microscope (SEM) images from the 
top facets and side edges by cleaving the samples or by Focused Ion Beam milling following the 
crystal planes.
\begin{figure}[b!]
  \includegraphics[width=8cm]{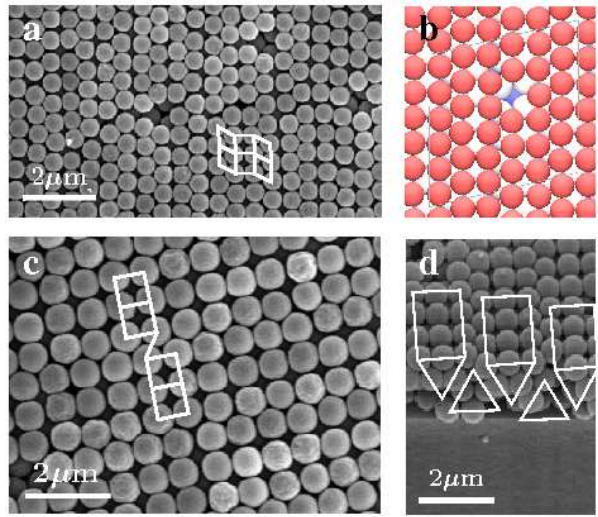}
  \caption{\label{fig:fig3} SEM micrographs of the prism phase found in this work: 
    $2P_A$ (a) and $3P^l_\square$ (c,d). A few particles were removed from the top layer upon detachment 
    of the covering plate allowing access to the structure in the layer below. A simulation snapshot, 
    where a particle was also removed (after the simulation), is shown in (b). White lines indicate 
    the symmetry of each phase (a,c) as well as the structure of the prism arrays 
    in the side view of $3P^l_\square$ (d). }
\end{figure}

Concentrating on the regime $3\square \to 3\triangle$, we found evidence of 
the adaptive $2P_A$ phase. Also, the $3P_{\square}^l$ phase has been observed
for larger plate separation distances. As an example,
SEM images of $2P_A$ and $3P_{\square}^l$ are shown in Fig.~\ref{fig:fig3}
along with a simulation snapshot for comparison.

Experiments on colloidal systems, such as ours, are necessarily performed at finite pressure.
In order to investigate the stability of the new prism phases away from close packing,
we performed Monte Carlo simulations at a fixed lateral pressure $P_{l}=-H^{-1}\partial F/\partial A$,
where $F$ is the free energy and $A$ denotes the area of the system. This definition of pressure is such 
that it approaches the bulk pressure as $H$ increases. 
The discovery of new crystal phases in this and previous theoretical works at infinite pressure after
the previous simulation work that addressed the stability at finite pressure begs 
the question how stable  these phases are  at a high, but finite pressure \cite{NielabaPRE2009}.
We simulated the system at a high pressure $P_l\sigma^3/k_BT=40$, for which the system would 
equilibrate within a reasonable time (for comparison the bulk crystallization pressure is 
$P\sigma^3/k_BT=11.56$~\cite{ZykovaTiman_coex_HS_AO}).
\begin{figure}[t!]
  \includegraphics[width=8cm]{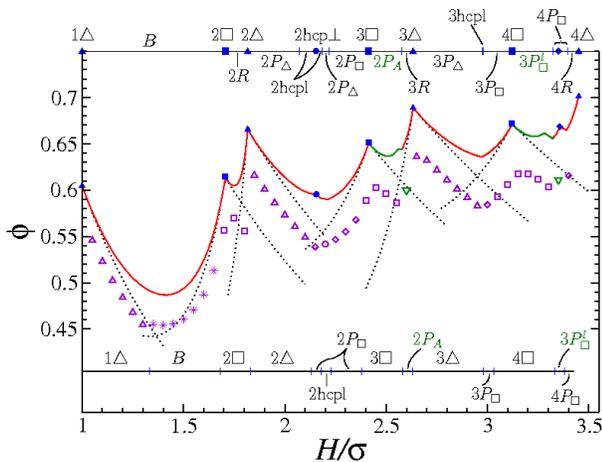} 
  \caption{\label{fig:fig4} Isobaric packing fractions $\phi$, as measured in simulations at fixed lateral pressure 
    $P_l\sigma^3/k_BT=40$ for confined hard spheres, versus dimensionless separation $H/\sigma$ (empty symbols) 
    compared to the theoretical results of Fig.~\ref{fig:fig2} (lines). Each type of empty symbol denotes a different 
    phase. The observed phases are indicated by the symbols on the horizontal lines at the top (theory) and 
    bottom (simulation) of the graph. 
  }
\end{figure}
The success of cell theory--effectively a single-particle theory--at high densities indicates that phase 
behavior at high pressures can be accurately modeled using relatively small systems. 
Our variable-shape simulation box contained $m\times m\times n$ particles, where $n$ is the number of layers and
$4 \leq m \leq 8$ ~\cite{suppinfo}. In Fig.\ \ref{fig:fig4}, we compare finite-pressure simulation data to theoretical 
results at infinite pressure. 
We clearly see that the packing fractions in both cases feature a qualitatively similar course. 
However, some phases vanish for finite pressure as this regime is dominated by broadened stability 
regimes of $n\triangle$ and $n\square$ phases. 
In detail, the $2R$, $3R$, $2\mathrm{hcp}\bot$, $3\mathrm{hcpl}$, $2P_{\triangle}$ and $3P_{\triangle}$ 
phases are not found for the finite pressure and accuracy $H/\sigma \pm 0.025$ chosen in the 
simulations. As can be further seen, the adaptive prism phase $2P_A$ and $3P^l_{\square}$ found in this work are 
stable at this pressure and, therefore, also at all higher pressures. 

These simulations help explain the absence of the triangular prism phase in the experiments 
(see \cite{Manzano2009}). We also performed simulations with the triangular 
prism phases as initial configuration. At the values for $H$ where the triangular prism phase has the 
highest density of all possible phases, the $nP_\triangle$ phase appears to consist of
$n$ only slightly distorted hexagonal layers. At finite pressure, the small distortions can 
quickly disappear and a regular triangular crystal can be formed. This is a typical scenario 
for crystal--crystal transitions for hard particles, where the close packed crystal phase transforms 
into a higher--symmetry crystal with a slightly lower density, but a greater entropy, 
as the pressure is decreased sufficiently. At larger values of $H$ than those investigated here, 
the triangular prisms are significantly enhanced and do not transform so easily to triangular 
crystals (cf.\ \cite{Fortini}). Preliminary simulations show that indeed stable 
triangular prim phases can be found for larger values for $H$ (see \cite{suppinfo}).

In conclusion, we explored the close-packed structures of hard spheres confined 
between hard plates in a broad range of plate separations by combining theory, experiment 
and simulation. We identified adaptive prism phases with rhombic symmetry which pack 
densest in certain ranges of the slit width.
An adaptive prism phase optimizes packing by adjusting its base symmetry flexibly to the slit width.
Also, we showed a high persistence of these adaptive prism phases at finite, but large pressure using 
experiments and simulations. We anticipate that adaptive prism phase will play a key role
for even higher plate distances, $H/\sigma > 3.5$, as ideal interpolating close-packed building blocks.

The adaptive prism phases found here offer new opportunities for several 
applications. For example, the reported structures possess pronounced symmetry directions whose 
alignment can be internally controlled by the slit height instead of using external fields 
(eg.\ electric fields, cf.\ \cite{Bartlett2011}). As a consequence, these phases can serve as switchable 
materials. 
Furthermore, we expect an unusual and anisotropic dynamical response of the multilayered prism 
phases upon shear \cite{CohenPRL2004} with possibly molten grain boundaries which can be exploited to tune 
the rheological properties of thin crystalline sheets. 
Finally, by varying $H$, it is possible to tune the whole complex cascade of close-packed structures. 
This may be of importance to fabricate nano-sieves or porous membranes \cite{Goedel} in a controlled way.

We thank Elvira Bonet, Mois\'{e}s Gar\'{i}n, and Kevin Mutch for helpful discussions. This work was partially supported by the 
DFG within the SFB TR6 (project D1), and by the Spanish CICyT projects, FIS2009-07812 and PROMETEO/2010/043.
F. R.-M. acknowledges the support from the EU Marie Curie project APPCOPTOR-275150 (FP7-PEOPLE-2010-IEF). 

\bibliographystyle{apsrev4-1}

\end{document}